# Improving the Hole Picture: Towards a Consensus on the Mechanism of Nuclear Transport

## David Cowburn[1] & Michael Rout[2]

1. Depts of Biochemistry and Systems & Computational Biology, Albert Einstein College of Medicine, Bronx, NY 10461, USA
2. Laboratory of Cellular and Structural Biology, The Rockefeller University, New York, NY 10065, USA

**ABSTRACT**

Nuclear pore complexes (NPCs) mediate the exchange of materials between the nucleoplasm and cytoplasm, playing a key role in the separation of nucleic acids and proteins into their required compartments.  The static structure of the NPC is relatively well defined by recent cryo EM and other studies. The functional roles of dynamic components in the pore of the NPC, phenylalanyl-glycyl (FG) repeat rich nucleoporins, is less clear because of our limited understanding of highly dynamic protein systems. These proteins form a restrained concentrate which interacts with and concentrates nuclear transport factors (NTRs) to provide facilitated nucleocytoplasmic transport of cargoes. Very rapid exchange among FG repeats and NTRs supports extremely fast facilitated transport, close to the rate of macromolecular diffusion in cytoplasm, while complexes without specific interactions are entropically excluded, though details on several aspects of the transport mechanism and FG repeat behaviors remain to be resolved. However, as discussed here, new technical approaches combined with more advanced modeling methods will likely provide an improved dynamic description of NPC transport, potentially at the atomic level in the near future.  Such advances are likely to be of major benefit in comprehending the roles the malfunctioning NPC plays in cancer, aging, viral diseases, and neurodegeneration.

**INTRODUCTION: Structure of a Behemoth**

Nuclear pore complexes (NPCs) are among the largest macromolecular assemblies in a eukaryotic cell. NPCs sit in the double-layered nuclear envelope (NE), which provides a barrier separating the nucleoplasm from the cytoplasm. There, NPCs form a platform for the organization of numerous nuclear functions, and critically, act as the sole mediators of macromolecular trafficking into and out of the nucleus. Malfunction of the NPC or its components are linked to many disease states (1-3). Steady progress has been made in discerning the fine structure of the NPC in yeast and vertebrates, resulting in recent relatively good consensus maps of the NPC's architectural principles (Fig. 1); each NPC is an octagonally near symmetric cylindrical assembly some 100 nm across and 50 - 100 MDa in mass (depending on species), comprised of ~500 proteins (termed Nups) representing ~30 different types that have been fully cataloged for both yeast and vertebrates (reviewed in e.g. (4-6)).

In contrast to this structural understanding, significant areas concerning the dynamic mechanism of the transporting NPC remain undefined. Overall, the dynamic interactions that mediate and regulate transport span thirteen orders of magnitude in time, from picoseconds to 10s of seconds. As we need to comprehend this entire scale, no one approach is fully sufficient to give us a complete and unbiased view (7, 8). Here, we discuss why much of the NPC's transport mechanism has proven so refractory to mainstream structural approaches, and how this has led to significant confusion both inside and outside the field, as well as contradictory models representing the full complexity of the structure, dynamics, and biology of the NPC.  On the positive side, recent ingenious and orthogonal investigations from multiple groups have begun to overcome prior limitations, and development of exciting new methods is likely to provide major new insights, that already hint that the transport mechanism is perhaps more complex and surprising than previously anticipated.

**MAIN BODY**

**The Road from Static Representation to Functional Understanding**

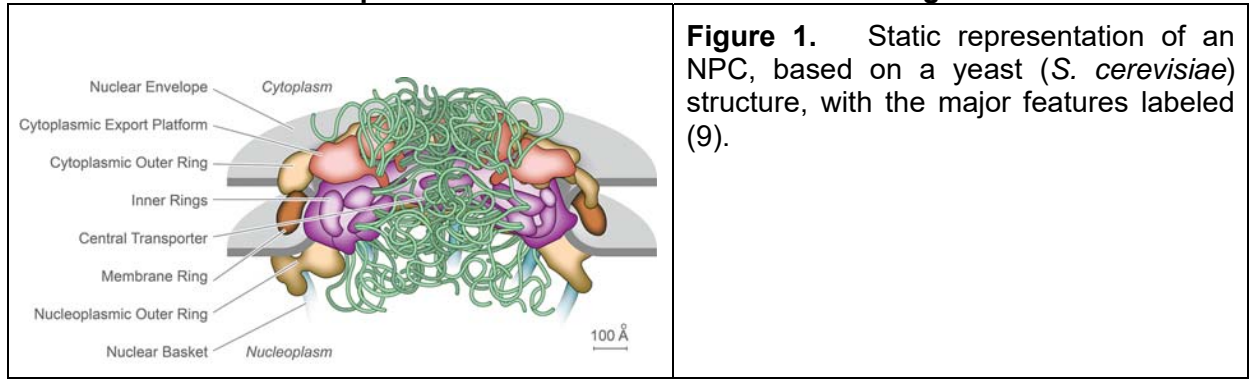

**Figure 1.**   Static representation of an NPC, based on a yeast (*S. cerevisiae*) structure, with the major features labeled (9).

In the nuclear envelope (NE), NPCs have two critical properties. First, they are the stationary phase of nuclear transport, mediating the mobile phase, comprising the bi-directional traffic of import of proteins to and export of RNAs from the nucleus (10, 11).

Much of transport across the NPC is mediated by multiple members of the karyopherin (Kap) family of nuclear transport receptors (NTRs), at rates approaching 1000 molecules / NPC / sec ((12, 13) and references therein). Import-Kaps (importins) transport cargos into the nucleus while export-Kaps (exportins) ferry cargos out of the nucleus. Protein cargos are targeted for transport by having a nuclear localization signal (NLS) or export signal (NES). NLSs/NESs bind Kaps, which, in turn, translocate through the NPC, after which the Kap-cargo complex dissociates in its target compartment; their transport directionality is controlled by the nucleotide state of the GTPase Ran, shuttled across the NPC by its dedicated transporter NTF2/p10, a representative of the other major NTR family whose other members, Mex67/NXF1 - Mtr2/NXT1, mediate the export of mRNAs. Other RNAs are exported by cognate Kaps, either directly or via adaptor proteins (and in the case of the 60S pre-ribosomal subunit, also utilizing Mex67/Mtr2) (reviewed in (14)). While small molecules such as metabolites and ions can freely diffuse across the NPC, macromolecules not associated with NTRs cannot pass as efficiently through the NPC, which thus functions as a selective barrier in the CT, leading to the distinction between fast facilitated diffusion (i.e., NTRs/cargoes) and slow or negligible passive diffusion of other macromolecules. While it was previously thought that there was something of a hard upper limit of ~40 kDa or ~9 nm radius for this passive diffusion, we now know that there is a power relationship between macromolecular size and the efficiency of its exclusion from the CT (15, 16); which  means that   few macromolecules access the nucleus through passive diffusion (17)

**The Transport Mechanism is Driven by Dynamics and Disorder**

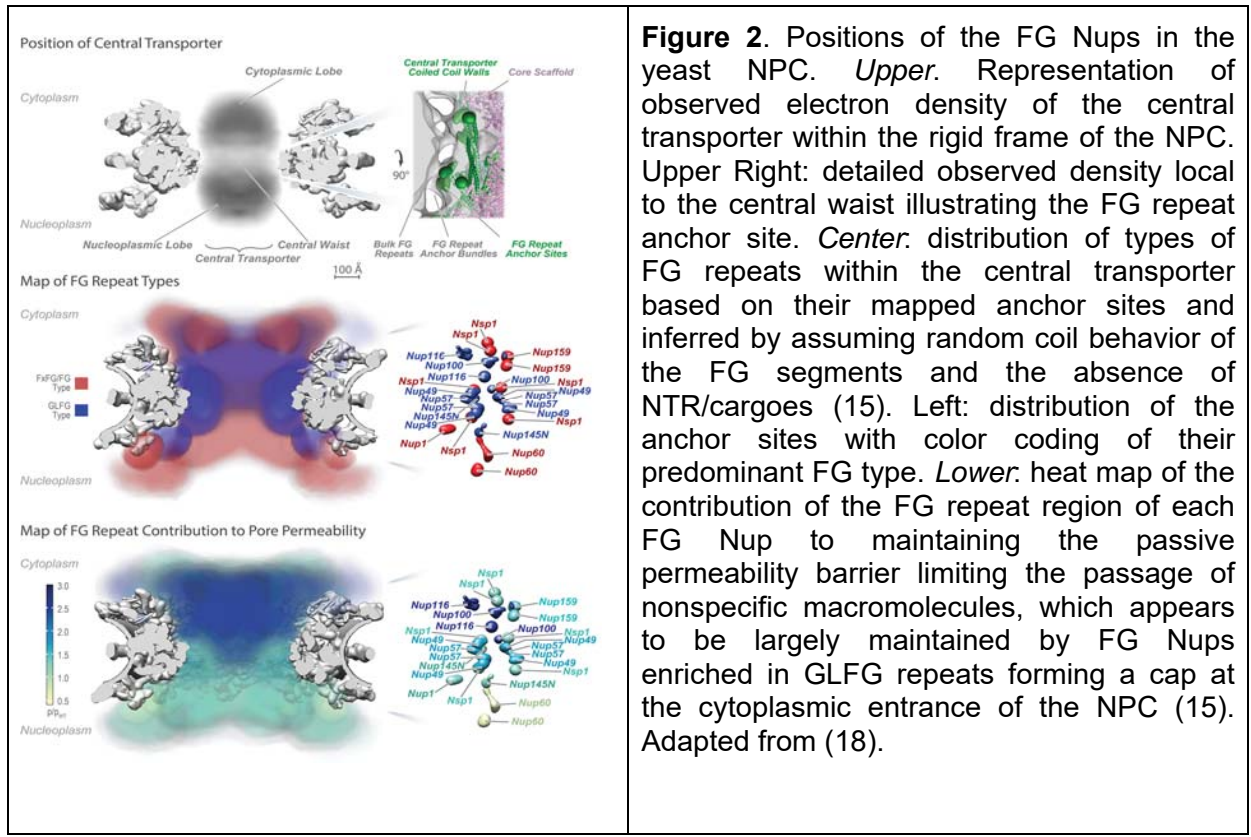

**Figure 2**. Positions of the FG Nups in the yeast NPC. *Upper*. Representation of observed electron density of the central transporter within the rigid frame of the NPC. Upper Right: detailed observed density local to the central waist illustrating the FG repeat anchor site. *Center*: distribution of types of FG repeats within the central transporter based on their mapped anchor sites and inferred by assuming random coil behavior of the FG segments and the absence of NTR/cargoes (15). Left: distribution of the anchor sites with color coding of their predominant FG type. *Lower*: heat map of the contribution of the FG repeat region of each FG Nup to maintaining the passive permeability barrier limiting the passage of nonspecific macromolecules, which appears to be largely maintained by FG Nups enriched in GLFG repeats forming a cap at the cytoplasmic entrance of the NPC (15). Adapted from (18).

It was originally proposed that the NPC's selective barrier might utilize mechanoenzymes, either by an iris-like gate or motor-driven translocation at the NPC through a permeability barrier of some kind, or that the translocation of NTRs across the NPC was propelled directly by cycles of Ran GTP hydrolysis (19-21). However, it has been shown that nucleotide hydrolysis is not required for the translocation step across the NPC. Instead, diffusion seemed to be key to trafficking, involving a restriction of passive diffusion and promotion of selective diffusion within the CT. An important clue to the transport mechanism came when it was shown that the CT is lined with proteins termed FG Nups (22). FG Nups are so-called because they contain large intrinsically disordered regions (IDRs) that carry many Phe-Gly (FG) repeats, each separated by ~20 residues of predominantly hydrophilic linkers (Fig. 2,3) (11).

Approximately one-third of all Nups contain these regions, which are in the volume of the CT. Changing the dynamic and disordered states of such IDRs is entropically unfavorable, such that any macromolecule attempting to enter their space or push them aside experiences an "entropic repulsion" effect (23). Crucially, it is these FG repeats that were also shown to interact with multiple cognate sites on each NTR and so specifically facilitate its passage across the NPC (24-26). Based on this information, a "virtual gating" model was proposed in which dynamic multivalent interactions of NTRs with these FG repeats would provide sufficient avidity to allow their rapid passage across the CT by overcoming entropic repulsion effects of the same IDR regions that otherwise exclude the passage of non-binding, non-specific macromolecules (22, 27). Indeed, it now seems evident that the mechanism for facilitated transport in the CT must include three features: first, the rate of facilitated transport across the NPC is similar to free diffusion within cells (28), so the internal mechanism of facilitated selection must be extraordinarily rapid; second, to maintain facilitated selection, the ratio of concentration of NTRs/cargoes to passive molecules (non-NTRs) within the CT must exceed that external to the CT, i, e, NTRs/cargoes must be relatively concentrated in the CT by interaction with it (9, 29-31); and third, to be consistent with the first and second points, the simplest mechanism of inhibition of non-NTR transport is entropic exclusion (10, 27, 32, 33), although exactly how that mechanism plays out in the NPC is still unclear (below). We will now address how these features are produced by the NPC.

**Near to the Madding Crowd: Complexity and Crowding within the Central Transporter Generates Specificity in Transport**

Frustratingly, the fact that the CT's component materials are either intrinsically disordered (FG repeats) or extremely heterogeneous (NTRs and their cargoes) has made structure-function studies of it extraordinarily challenging, such that it has been described as "structurally elusive and mechanistically controversial" (7). Collectively, the anchor sites for the FG repeats in the walls of the CT direct them towards the CT's central axis to generate a highly concentrated and dynamic FG repeat milieu; the anchor sites for most FG repeats are clustered, so that they emanate as bundles near the walls of the CT which can be visualized by electron microscopy, and then merge into a cloud near the CT's axis (9, 18). (Fig. 2). This generates two organizational features: firstly, the regions of FGs in the bundles near the CT's wall are more diffusionally restricted, as has been indicated in vivo (34, 35); and secondly, different kinds of FG repeat (termed "flavors") are at specific positions in the CT's volume (Fig. 2). These flavors can be divided into two broad classes based on an approximate consensus of their Phe-containing repeats and the amino acid composition of the repeat spacers: one of Phe-X-Phe-Gly (FXFG) - like repeats (where X is usually a small hydrophilic amino acid) with hydrophilic spacers often carrying some charged amino acids (Asp, Glu or Lys), and the other of Gly-Leu-Phe-Gly (GLFG) or Phe-Gly (FG) - like consensus repeats spaced by hydrophilic segments of low charge, although there is considerable variation between FG Nup flavors and between the same FG Nup homologs of different species (11). One possible reason for such flavor varieties is that they confer different biophysical properties to specific positions in the CT. Thus, the role of differences in charged residues in the spacer regions has been suggested to be of significance in the CT by repulsion between like-charged sequences (36) and partitioning of charged and less charged sequences resulting in a permeability barrier (37). Compelling in vivo data also point to the idea that these different flavors and their locations in the CT delineate specific pathways for subsets of NTRs through the NPC (38-45) (46), e.g. it is clear membrane proteins can be actively transported through the NPC in a route distinct from those of soluble cargoes (47, 48).

Surprisingly, despite their high concentration the FG repeats do not comprise the majority of the CT - rather, well over three-quarters of the CT at any moment is made of a constant flux of NTRs and their cargoes (see below) (9, 18, 29, 31). Their sizes vary widely up to many tens of megadaltons, with ribonucleoprotein / mRNA (termed mRNP) cargoes and ribosomal subunit precursors being a significant fraction of the total transport flux(18, 49). Recently experiments on large cargoes indicated that the increased free energy cost of inserting a large cargo into the dense FG Nup barrier is compensated by the binding to FG Nups via more NTRs per cargo (50), and the very largest cargoes may require expansion of the NPC in some fashion (9, 51-53). This enormous preponderance of NTRs and their cargoes in the CT is the elephant in the room; earlier models have concentrated solely on the roles and states of the FG Nups (below), but in nature, FG repeats in the NPC always exist in the presence of a considerable molar excess of NTRs, meaning that experiments that reconstitute FG repeats in the absence of NTRs could well be examining unnatural states.

We are thus faced with accounting for how three quarters or more of the CT's mass contributes to its transport behavior, and this realization has led to a recent shift in focus, onto the interplay between FG repeats and NTRs, rather than just the FG repeats alone (54). Crucially, this high concentration of NTRs and cargoes in the NPC, all specifically enriching around the FG repeats, can outcompete and so inhibit nonspecific macromolecular exchange which cannot interact with FG repeats (31, 55, 56). Moreover, there is mounting evidence for a slowly exchanging pool of NTRs being maintained at the CT (31, 57, 58). Such observations have led to a "Kap-Centric" model, wherein there is a slower exchanging pool of NTRs that are key players in modulating and maintaining the NPC's barrier to non-specific macromolecular exchange (31, 59, 60). In summary, "the FG Nups are necessary but insufficient for NPC barrier function. NTRs constitute integral constituents of the NPC whose barrier, transport, and cargo release functionalities establish a continuum under a mechanism of Kap-centric control" (31).

**Fast and Furious: Active Macromolecular Transport Across the NPC through NTR-FG Repeat Interactions**

A defining characteristic of nuclear transport is the tremendous rates at which each NPC can bidirectionally transport an astonishing variety of cargoes. The family of NTRs is quite large and still may not be fully defined. Remarkably, dozens of different NTRs each mediate separate but often overlapping transport pathways for specific classes of cargoes across the NPC. This multitude of FG interaction sites on each NTR (above; see also Fig. 3) presents the question of how rapid transport could avoid slowing by avidity. Generally, the time scales and energies of interactions between FG repeats and NTRs appear very rapid, and so difficult to measure either in vivo or in vitro (8, 61). The specificity of the interactions is clearly linked strongly to the phenylalanyl side chain of FG Nups as revealed by crystallography, NMR, MD simulations, and other methods(62-64). From solution methods (61, 65), and consistent with in situ high speed AFM (66), the time scale of NTR-FG interactions is likely of the order of microseconds, so that e.g. "weak and ultrafast multivalent Kap–FG interactions allow the Kap–cargo complexes to translocate in a fast and selective manner" (7). Atom scale molecular dynamics, supported by NMR data, indicated that the fast exchange between NTRs and individual FG motifs may rely on a sliding-and-exchange mechanism (64), indicating that FG motifs slide on the ample grooves that form NTRs' binding pockets. This anisotropic sliding may in turn enable fast exchange and rapid facilitated diffusion, such that interacting FG repeats and NTRs exchange particularly rapidly compared with other protein-protein interactions of similar affinity, allowing for the remarkable transport rates observed experimentally (Fig. 3).

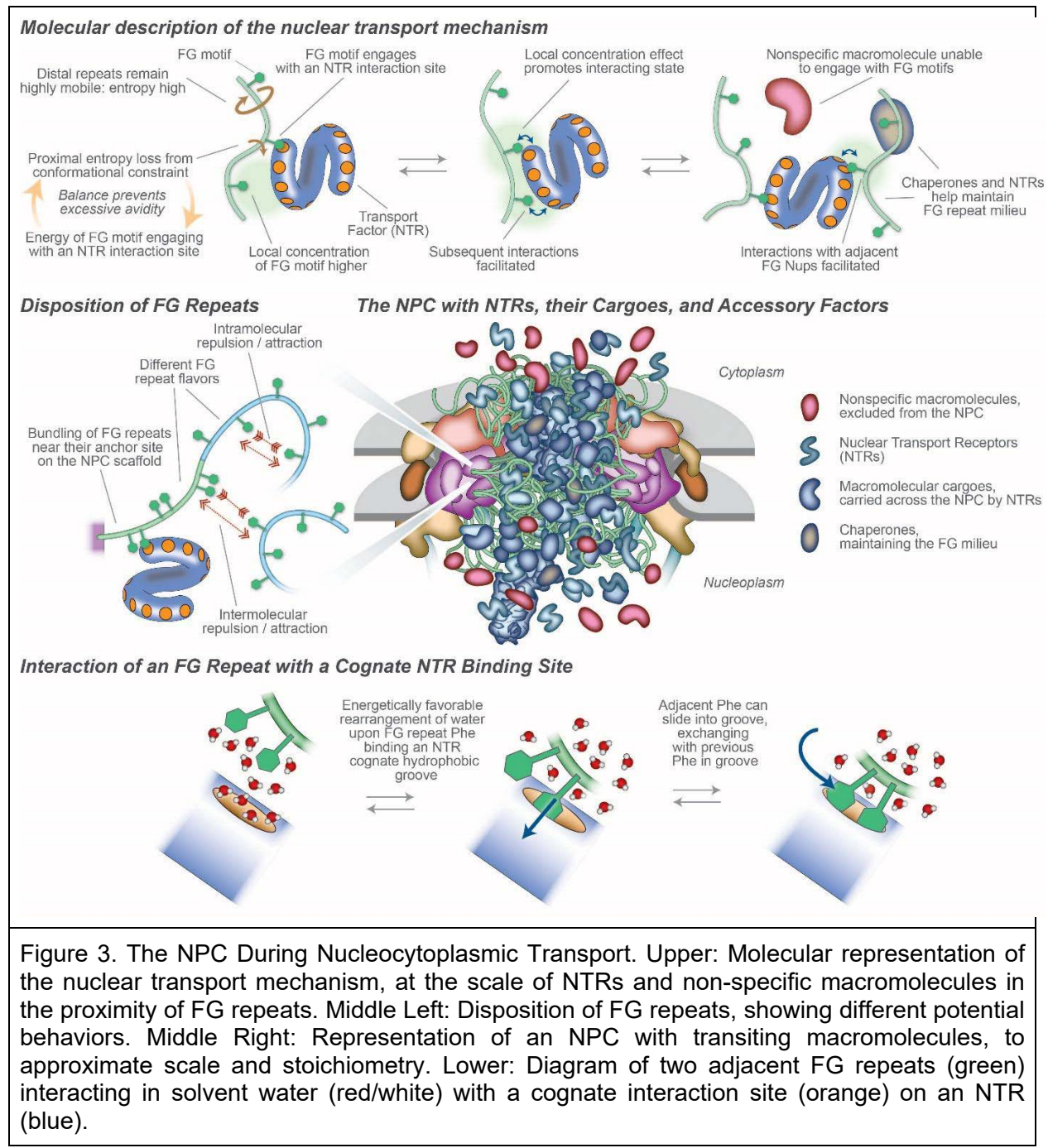

Figure 3. The NPC During Nucleocytoplasmic Transport. Upper: Molecular representation of the nuclear transport mechanism, at the scale of NTRs and non-specific macromolecules in the proximity of FG repeats. Middle Left: Disposition of FG repeats, showing different potential behaviors. Middle Right: Representation of an NPC with transiting macromolecules, to approximate scale and stoichiometry. Lower: Diagram of two adjacent FG repeats (green) interacting in solvent water (red/white) with a cognate interaction site (orange) on an NTR (blue).

The emerging picture is that NTRs (with or without cargoes) can transiently skip between FG sites on FG Nups with a low interaction enthalpy dependent on the local concentration of FG Nups and limited in avidity by the entropic motions of the FG Nups' IDP character (67), while passive diffusion is limited by entropic exclusion, with minimal benefit from interactions with the FG sites. Where the energy for interaction of an F with an NTR pocket comes from is still unclear, but as well as direct amino acid interactions, it seems likely that rearrangement of water molecules from around the hydrophobic F residue and pocket may play a part (Fig. 3). High avidity is avoided as more and more FG motifs interact at an NTR's multiple pockets via an application of the "virtual gate" model, in which the modest enthalpy of a single interaction of an FG with a given cognate NTR binding pocket (~ -7 kCal/M) is offset by the entropic cost of constraining the FG repeat plus spacer to the NTR's vicinity (~ 3 kCal/M). More interactions yield non-linearly more interaction enthalpy, but that is balanced by a similar increase in entropic cost, this being only slowly augmented by a local concentration effect on enthalpy limited to 4-6 additional FG sites providing a local concentration effect (67) (Fig. 3). One analogy is that the FG Nups form a "cloud" of rapidly diffusing phenylalanine ligands that are constrained to the vicinity of the CT, acting like a "solvent" for the NTR/cargoes with superfast, transient, and weak interactions, although unlike normal solvents the interaction sites are linked (Fig. 3). Additional complexity is added by the specificity of interactions between classes of FG Nups and specific NTRs, and by potential interactions ("cohesion") among the components.

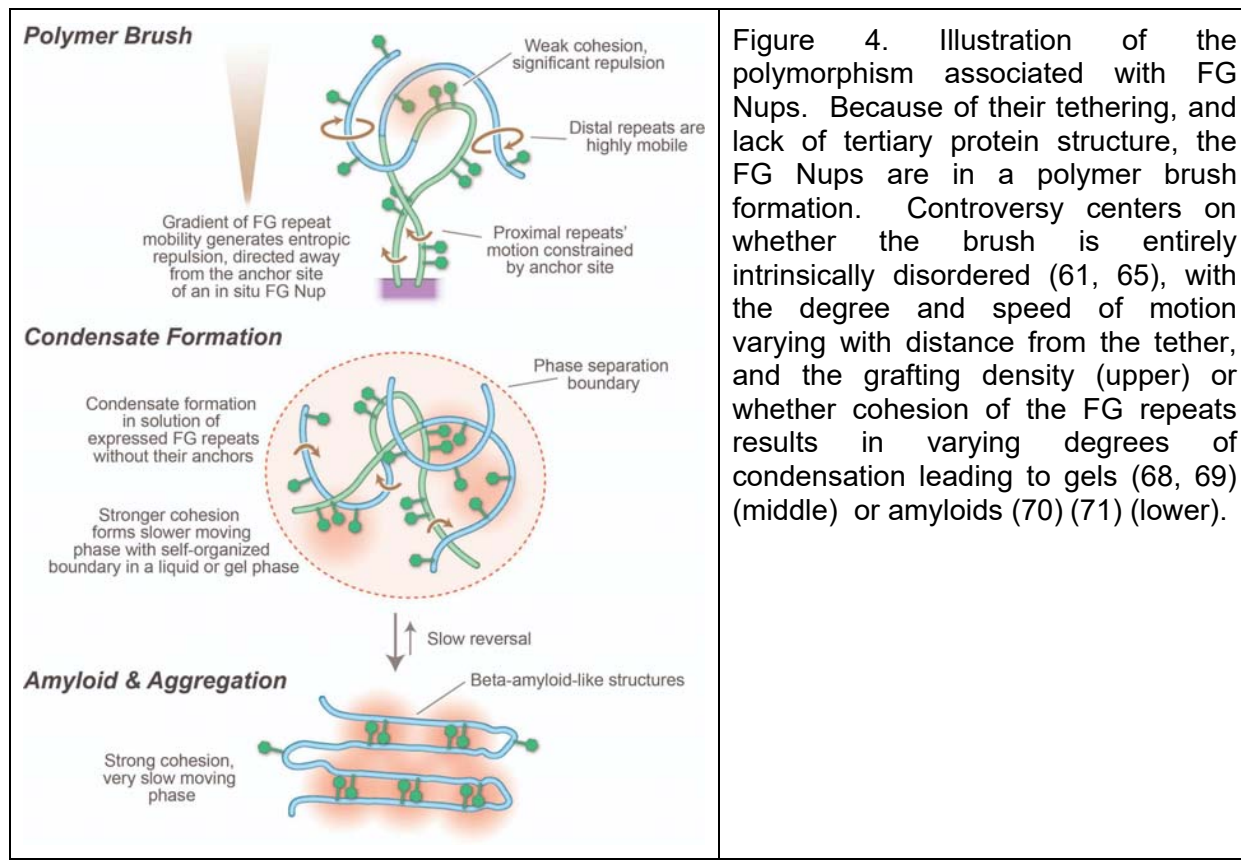

Figure 4. Illustration of the polymorphism associated with FG Nups. Because of their tethering, and lack of tertiary protein structure, the FG Nups are in a polymer brush formation. Controversy centers on whether the brush is entirely intrinsically disordered (61, 65), with the degree and speed of motion varying with distance from the tether, and the grafting density (upper) or whether cohesion of the FG repeats results in varying degrees of condensation leading to gels (68, 69) (middle) or amyloids (70) (71) (lower).

**FG Repeat Biophysical Behaviors: The Quick and the Dead?**

What kind of "solvent" might the FG repeats form in the CT, and in addition to the NTRs' role, how might its physical states contribute to exclusion of non-specific materials? Biologically, solvents are normally liquids, but the FG repeats cannot form a liquid, because the FG repeats are not freely mobile but rather tethered. As the nature of the barrier to diffusion is hard to discern in NPCs in a cell, various approaches to mimicking the roles of FG Nups have been used. Significantly, selective transport can be mimicked by nanopores of defined composition with FG Nups incorporated using synthetic membranes (55, 60, 72-75), which demonstrates that the essential phenomenon of selectivity can be reconstructed ex vivo, although providing little information about the state in the CT. However, when isolated as free proteins in vitro,  FG repeats can assume a range of polymorphisms, from simple intrinsically disordered proteins (IDPs) in solution, to polymer brushes tethered to a surface, to different classes of condensate: liquid-liquid phase separations(68), hydrogels (76), prions(77, 78), or rigid amyloid-like gels (70) (Fig 4; BOX 1). These states span a substantial range of viscoelastic properties (see BOX 1)(79). Moreover, all display in vitro at least some of the characteristics also seen in situ, though none have the performance of native NPCs (75). This range of FG repeat regions' states and behaviors is reflected in different proposed models for their function in the CT in vivo (reviewed in (37)). But this is putting the cart before the horse - all these states can be maneuvered to display certain transport characteristics does not address what state the FG repeats form in the NPC and so which is the actual state, and speed of FG repeat conformational fluctuations, in vivo, remain a subject of investigation and debate.

When IDPs are tethered to a surface, as they are in the CT, they will form a polymer brush (BOX 1), evidence for which is seen for FG repeats in vivo (80) and in isolated NPCs (9, 18) (Figs. 3,4). Inclusion of any other macromolecule in the brush is entropically unfavorable as it restricts the brush's freedom of motion, resulting in an force pushing away from the brush; in the CT, this would result in a kinetic barrier excluding non-specific macromolecules from the pore, this forces scaling with the size of the macromolecule (15, 22, 27). FG repeats grafted to surfaces (reviewed in (45)) and pores carrying walls grafted with disordered polymers generate such an effective kinetic barrier (81). While NTRs would also experience (and provide; see above) entropic exclusion, their avidity to the same FG repeats would offset this entropic exclusion and so allow their rapid passage across the NPC (22, 27), as quantitatively demonstrated for FG repeats and NTRs in vitro (67).

Increasing condensation from freely soluble to rigid gels in these materials is referred to in the field as the result of 'cohesiveness' (Fig.4). Note that cohesiveness is used as an omnibus term covering all weak interactions associated with compactness within a single chain, interactions between chains of the same type (homotypic) and interactions with other FG Nups in the vicinity (heterotypic). Based on observations of apparent radius of gyration (82), the GLFG flavor of FG repeats (above) has generally been assigned a higher degree of cohesiveness. In solution this is reflected by isolated solutions of GLFG repeats being able to display the complete range of polymorphisms (above) (68, 70, 71, 83). There is no direct evidence testing which of these states or combination of states is present in the NPC, even though the suggestion that a high degree of condensation (or cohesion) plays a functional role has been widely propagated (69, 70, 76, 83-85) (86). The nature of the stabilizing cross links in such gels is also unclear, as is how NTRs reversibly dissolve these links. While the suggestion that Phe-Phe interactions may provide such links to form a "hydrophobic gel" (28, 69), there is no evidence for significant Phe-Phe interactions in solution or in gel states by NMR (68, 70, 87, 88) or in solution by SANS (89), although such weak interactions are inferred in amyloids of Nup98 by cryo EM (71).

With the advent of liquid-liquid phase separation (LLPS) in cell biology (90, 91), the idea was raised that FG repeats, too, forming an LLPS. Indeed, under various conditions, soluble FG repeats can be induced to form LLPSs in vitro (85, 92-94). The debate now surrounds: which properties of LLPS the central transporter's FG repeats actually possess? Certainly, LLPSs have two characteristics that FG repeats in the NPC lack, namely, the polymers comprising LLPSs are untethered, and they are defined by their self-organized surface tension (95-97) (BOX 1).

| |
|---|
| **Box 1. Definitions of Different Potential FG Repeat States** |
| *Polymer brush* - "Polymer brushes are long-chain polymer molecules attached by one end to a surface or interface by some means, with a density of attachment points high enough so that the chains are obliged to stretch away from the interface, sometimes much farther than the typical unstretched size of a chain." (98). Therefore, the CT by definition contains polymer brushes, though additional properties of state may be displayed. |
| *Condensate* - dynamic and reversible assemblies of molecules that can dissolve and be reused to perform their function (99). The usual implication is that the density of a condensate is the result of weak intermolecular forces between the components, and the condensate has significant displacement of solvent from itself. A condensate may be a component of a liquid-liquid separated phase or may be part of a complex structure with components not able to form liquid phases e.g., chromatin (100). |
| *Gel* - material with elastic properties, usually with significant permeability to solvent through the gel itself exhibits no steady-state flow and is usually crosslinked. Gels are typically the result of condensates at high concentration (101). For the NPC, it usually describes in vitro formation from GLFG rich FG repeats, and is ascribed without direct evidence to possible function in the CT. |
| *Liquid-liquid phase separation* - "Phase separation describes the process by which a well-mixed solution of components de-mixes into two or more coexisting phases with uniform properties. In the simple case of liquid–liquid phase separation, a liquid solution de-mixes into two liquid phases, one dense phase and one dilute phase." (99). In LLPSs it is solely the surface tension from selective cohesion of its freely mobile components that defines a concentrated compartment (96). |
| *Viscoelastic complex* - As a result of the formation of some of the above items, the properties of the complexes may be highly variable from essentially fluid, responding to changes of pressure by diffusion, to more gel-like or solid properties, which are elastic, deforming to changes of pressure (102). |

Overall, the functional role in vivo for changes of state of FG repeats involving LLPS, gels or amyloids is entirely unknown and may even be unrelated to their function in nuclear transport. This issue of structural pleomorphism and its functional implications is a major challenge generally to the structural biology of amyloid and gel-like systems, complicating in vitro reconstitution e.g. (103). Moreover, CT assembly is likely tightly regulated, to balance functional activities against the aging and aggregation that has been implicated in leading to amyloid like disease states (99). Indeed, there appear to be active balance and maintenance systems that limit FG repeat condensation in vivo. Thus, the observation that removal of highly condensed FG Nups may be facilitated by chaperones (104, 105) interacting with FG repeats, suggesting that chaperone-like activity may play a role in assembling and maintaining the NPC. Moreover, other work indicates that NTRs contribute actively to preventing aggregation of FG repeats (106), and FG Nup condensate puncta outside the NPC appear to be transient non-essential and even toxic condensates that are absent in healthy cells (107) (also reviewed in (108)), and the more aggregated forms have been associated with disease states e.g. in neurodegeneration (109-111).

Taking all this together, it seems reasonable to suggest that the polymer brush FG repeats of the CT form a "restrained concentrate", where a concentrating effect of FG repeats is achieved by the anchoring of FG Nups in the CT walls, and the density and selective transport behavior of the CT contents results mainly from this constraint plus the recruitment of NTRs/cargoes (above), without necessarily having a *specific* requirement for internal cohesion necessary to form a gel. This view is most consistent with solution studies of different isolated FG repeat flavors which show a picture of highly mobile, minimally cohesive IDPs (61, 65, 67, 112) with fast, low affinity interactions with NTRs and with very high mobility of tagged FG repeat regions in vivo (34, 53, 80, 113). Notable in this regard, the Lim lab has also contributed key observations that the central portion of the in situ CT is dynamic using high speed atomic force microscopy, definitively establishing the movement of the center of the NPC in the 100 ms and faster range and showing that intermingling FG Nups do not appear to cohere into a highly crosslinked meshwork (35, 66). Some form of weak cohesive forces must exist to some (currently ill-defined) degree; for example, the density and packing of the FG Nups and NTRs/cargoes are subject to the usual dynamic interactions including van der Waals attractive and repulsive forces, complementing specific FG/NTR interactions in which NTRs may bridge between different FG repeats, and other potential specific interactions (37, 114, 115). These forces may have some "tuning" role to play in adjusting the permeability and selectivity parameters of the CT.

**Pores in Action: A New View In Vivo**

Recent innovative approaches have begun to move the field away from the drawbacks of reliance on only in vitro data, to now garnering detailed nanoscale dynamic data on NTRs and FG repeats in the CTs of in situ or living NPCs. In particular, the examination of functional FG Nups by measuring fluorescence energy transfer between two neutral fluorophores placed at different positions along the length of FG repeats in the in vivo NPC provides a significant advance in our understanding of their dynamic structure (113). Pioneering advances in design of small amino acid fluorophores, insertion of multiple labeling sites in the appropriate genes, and measurement of the distance distribution of eighteen NUP98 segments is consistent with the in-NPC state being close to that of a random polymer in a 'good solvent', and is significantly different from the value observed in solution at low concentration that is consistent with a compacted state (as also seen by other methods) (68). More detailed analysis of the fluorescence lifetime decay using simulations is consistent with extremely rapid polymer motion, and suggestive of some shuffled packing of the FG repeats towards the periphery with a concentration of NTR/cargoes towards the center (113). Similar new approaches in situ and in vivo are on the horizon, promising direct observation of transport in action at the molecular level.

Current research reviewed has also concentrated on the function and mechanism of the CT. However, other recent work has pointed to additional factors that may play critical roles in the transport mechanism. These include: a potential role of numerous transport factors in forming the nucleocytoplasmic Ran gradient (116); accounting for the dilation of the NPC in diffusion control and in response to environmental changes (9, 52, 117, 118); and the existence of multiple and very distinct NPC isoforms, even within the same cell, that may have different transport roles (9, 52). The diversity of NTRs, FG repeats and even NPC isoforms also raises the question as to whether only one general mechanism is employed by the NPC in selective transport and it even seems possible that different NTRs may employ distinct mechanisms, perhaps in concert with specialized NPC isoforms.

**New Tools on the Block: Data Integration, Modeling and Simulation**

On their own, these multiple biophysical and cell biological observations are not readily merged into a simple hypothesis of how the CT permits rapid and selective diffusion. Recent modeling papers have focused on the underlying thermodynamic issues of enthalpic/entropic balance in FG Nup/NTR interactions, on the dynamics associated with diffusion, on the roles of cohesion in FG Nup self-interactions, and on the dynamic architecture of the CT and its role in providing specific pathways for diffusion. These both integrate the prior observations and suggest new hypotheses testable by experiment.

While their role in vivo is unclear, FG repeat gels are significant as a model for phase separation studies, and an underlying thermodynamic model was recently developed (83) rationalizing the observed increased stability of gels with temperature (68). The critical limit for gel formation was also calculated for different amino acid compositions using coarse grained modeling at the bead-per-residue level (119). The effects of cohesiveness on the selective permeability of in vitro FG repeat assemblies were simulated over a wide range of cohesiveness, showing that an increase in cohesiveness leads to decreasing permeability but that permeability may be enhanced with weak cohesiveness (120). Modeling of the FG Nups in the CT in the absence of transport factors suggested a heterogeneous diffusion barrier of several condensates formed by electrostatic pairing rather than FG-FG interactions (37). The role of the dynamic architecture of the CT was modeled and compared to experimental fluorescence anisotropy data by use of a bead equivalent of 4 nm resolution, and it was proposed that FG repeats are highly mobile and can reptate throughout the CT on timescales similar to experiment (121). The role of transient formation of voids permitting a size dependent permeation was analyzed (122) using the Onck force field (123). Regarding modeling passive transport, a Brownian dynamics simulation, with FG repeats represented as spring-like polymer beads and passive diffusing macromolecules as rigid spheres, suggested that the barrier to non-specific diffusion resulted largely from the highly dynamic FG repeats and entropic exclusion (15).

Several recent papers address more directly the modeling of FG repeat's interactions with NTRs. Based on experimental data for FG repeats, NTF2, and non-specific components' interactions, agent-based modeling (124) discriminated between binding models to discriminate multivalent cases (56). Crowding by different NTRs affecting interactions with FG Nups were proposed from coarse-grained classical density functional theory application with two residues per bead model for FG Nups, and spheres for NTRs, with the significant conclusion that at high NTR concentrations, there is increased flux (45) consistent with experimental data (125). Complementing two experimental works on biomimetic nanopores with separate FG Nups (60, 72) coarse grained modeling suggested that the NTR Kap95 forms a stable population bound to the CT periphery with fast transport proceeding in the FG-rich central channel.

A fair judgment is that these models are based on a wide range of assumptions/parameters, and that currently their results, in terms of describing details of the transport mechanism, are rather divergent. As the field integrates more data and refines models in the areas of siting of CT components, of parametrization of interactions energies of FG repeats and NTR/cargoes, of increased simulation times comparable to transit times in the NPC, and of comparison to more detailed tracking of individual NTR transits (e.g. (38)), there will hopefully be some convergence towards a realistic representation of the mechanisms of transport. While our knowledge of how FG repeats and NTR/cargoes interact in vivo is limited, there is however consensus that "stronger interactions and higher concentrations can block the transport. Importantly, accumulation of the transport proteins in the pore can also impede the translocation of inert molecules" (8). Similarly, there is agreement that FG repeat/NTR interactions are key to selective diffusion, while there remains controversy in the models about the nature and role of FG repeat cohesive interaction and their role in limiting passive diffusion. A key missing ingredient of current simulations is lack of incorporation of the "elephant" --  our updated knowledge of the high density of NTR/cargoes within the CT, and perhaps also limited appreciation of the difficulty of understanding of the whole NPC - whose size and detail may require a so-called paradigm shift of approach rather than the cumulative accretion of data (126)

**PERSPECTIVES**

*A view of the field's importance*: every year, more and more connections are discovered between NPC dysfunction and a range of serious, widespread and challenging human diseases; these include many cancers, neurodegenerative diseases, and a host of viral diseases including most recently SARS-CoV-2(127, 128). Designing effective therapeutics for such dysregulations depends upon the biomedical community gaining a detailed and comprehensive understanding of all the functionalities associated with NPCs, foremost among these being the mechanisms underlying nucleocytoplasmic transport.

*A summary of main areas of current thinking*: There is now consensus on many aspects of the mechanisms underlying nucleocytoplasmic transport. FG Nups anchored in the walls of CT generate a brush of intrinsically disordered FG repeat regions that form a high local concentration of FG motifs. NTRs, often carrying cargoes, cross by binding these FG motifs. The Interaction between NTRs and FG repeats leads to a high local concentration of NTRs in the CT, further strongly contributing to competitive exclusion of non-specific macromolecules. Much of the remaining controversy revolves around how non-specific macromolecules are further prevented from crossing the NPC, with possibilities ranging from FG repeats forming slow moving highly crosslinked gels, to weakly (or essentially non-) cohesive FG repeats being highly mobile and entropically excluding only non-interacting macromolecules while facilitating rapid transit of NTRs with cargoes.

*Potential future directions*: Resolving the remaining controversies, discussed above, is obviously a major priority. We must also come to grips with the astonishing diversity and pliability now being revealed in the NPC's architecture and mechanisms; that different NPCs have different compositions and so may specialize for different transport pathways, that NPCs can change shape in such a way that may modulate transport, and that different cargo types may take different paths across the NPC with different mechanistic details and at different times. Finally, as more mechanistic links between NPCs and diseases are understood, the potential for therapeutic intervention in nucleocytoplasmic transport will likely greatly increase.

**Acknowledgments:**  We are grateful to the Cowburn and Rout labs for discussions.  Supported by NIH grants R01 GM117212 (DC), P41 GM109824 and R01 GM112108 (MPR).

**References**

1.	Mettenleiter TC. Breaching the Barrier-The Nuclear Envelope in Virus Infection. J Mol Biol. 2016;428(10 Pt A):1949-61.
2.	Chandra B, Michmerhuizen NL, Shirnekhi HK, Tripathi S, Pioso BJ, Baggett DW, et al. Phase Separation Mediates NUP98 Fusion Oncoprotein Leukemic Transformation. Cancer Discov. 2022;12(4):1152-69.
3.	Spead O, Zaepfel BL, Rothstein JD. Nuclear Pore Dysfunction in Neurodegeneration. Neurotherapeutics. 2022;19(4):1050-60.
4.	Hampoelz B, Andres-Pons A, Kastritis P, Beck M. Structure and Assembly of the Nuclear Pore Complex. Annu Rev Biophys. 2019;48:515-36.
5.	Petrovic S, Mobbs GW, Bley CJ, Nie S, Patke A, Hoelz A. Structure and Function of the Nuclear Pore Complex. Cold Spring Harbor perspectives in biology. 2022;14(12).
6.	Fernandez-Martinez J, Rout MP. One Ring to Rule them All? Structural and Functional Diversity in the Nuclear Pore Complex. Trends in biochemical sciences. 2021;46(7):595-607.
7.	Huang K, Szleifer I. Modeling the nucleoporins that form the hairy pores. Biochem Soc Trans. 2020;48(4):1447-61.
8.	Hoogenboom BW, Hough LE, Lemke EA, Lim RYH, Onck PR, Zilman A. Physics of the Nuclear Pore Complex: Theory, Modeling and Experiment. Phys Rep. 2021;921:1-53.
9.	Akey CW, Singh D, Ouch C, Echeverria I, Nudelman I, Varberg JM, et al. Comprehensive structure and functional adaptations of the yeast nuclear pore complex. Cell. 2022;185(2):361-78 e25.
10.	Rout MP, Aitchison JD. The nuclear pore complex as a transport machine. J Biol Chem. 2001;276(20):16593-6.
11.	Heinss N, Sushkin M, Yu M, Lemke EA. Multifunctionality of F-rich nucleoporins. Biochem Soc Trans. 2020;48(6):2603-14.
12.	Wing CE, Fung HYJ, Chook YM. Karyopherin-mediated nucleocytoplasmic transport. Nature reviews Molecular cell biology. 2022;23(5):307-28.
13.	Kalita J, Kapinos LE, Lim RYH. On the asymmetric partitioning of nucleocytoplasmic transport - recent insights and open questions. J Cell Sci. 2021;134(7).
14.	Ashkenazy-Titelman A, Shav-Tal Y, Kehlenbach RH. Into the basket and beyond: the journey of mRNA through the nuclear pore complex. Biochem J. 2020;477(1):23-44.
15.	Timney BL, Raveh B, Mironska R, Trivedi JM, Kim SJ, Russel D, et al. Simple rules for passive diffusion through the nuclear pore complex. J Cell Biol. 2016;215(1):57-76.
16.	Popken P, Ghavami A, Onck PR, Poolman B, Veenhoff LM. Size-dependent leak of soluble and membrane proteins through the yeast nuclear pore complex. Mol Biol Cell. 2015;26(7):1386-94.
17.	Wuhr M, Guttler T, Peshkin L, McAlister GC, Sonnett M, Ishihara K, et al. The Nuclear Proteome of a Vertebrate. Current biology : CB. 2015;25(20):2663-71.
18.	Kim SJ, Fernandez-Martinez J, Nudelman I, Shi Y, Zhang W, Raveh B, et al. Integrative structure and functional anatomy of a nuclear pore complex. Nature. 2018;555(7697):475-82.

19. Akey CW. Visualization of transport-related configurations of the nuclear pore transporter. Biophys J. 1990;58(2):341-55.
20. Stewart M. Nuclear pore structure and function. Semin Cell Biol. 1992;3(4):267-77.
21. Ryan KJ, Wente SR. The nuclear pore complex: a protein machine bridging the nucleus and cytoplasm. Curr Opin Cell Biol. 2000;12(3):361-71.
22. Rout MP, Aitchison JD, Suprapto A, Hjertaas K, Zhao Y, Chait BT. The yeast nuclear pore complex: composition, architecture, and transport mechanism. J Cell Biol. 2000;148(4):635-51.
23. Bolthausen E, Deuschel JD, Zeitouni O. Entropic Repulsion of the Lattice Free-Field. Communications in Mathematical Physics. 1995;170(2):417-43.
24. Clarkson WD, Kent HM, Stewart M. Separate binding sites on nuclear transport factor 2 (NTF2) for GDP-Ran and the phenylalanine-rich repeat regions of nucleoporins p62 and Nsp1p. J Mol Biol. 1996;263(4):517-24.
25. Bayliss R, Ribbeck K, Akin D, Kent HM, Feldherr CM, Gorlich D, et al. Interaction between NTF2 and xFxFG-containing nucleoporins is required to mediate nuclear import of RanGDP. J Mol Biol. 1999;293(3):579-93.
26. Bayliss R, Corbett AH, Stewart M. The molecular mechanism of transport of macromolecules through nuclear pore complexes. Traffic. 2000;1(6):448-56.
27. Rout MP, Aitchison JD, Magnasco MO, Chait BT. Virtual gating and nuclear transport: the hole picture. Trends Cell Biol. 2003;13(12):622-8.
28. Ribbeck K, Gorlich D. Kinetic analysis of translocation through nuclear pore complexes. Embo J. 2001;20(6):1320-30.
29. Kapinos LE, Schoch RL, Wagner RS, Schleicher KD, Lim RY. Karyopherin-centric control of nuclear pores based on molecular occupancy and kinetic analysis of multivalent binding with FG nucleoporins. Biophys J. 2014;106(8):1751-62.
30. Lim RY, Huang B, Kapinos LE. How to operate a nuclear pore complex by Kap-centric control. Nucleus. 2015;6(5):366-72.
31. Kapinos LE, Huang B, Rencurel C, Lim RYH. Karyopherins regulate nuclear pore complex barrier and transport function. J Cell Biol. 2017;216(11):3609-24.
32. Matsuda A, Mofrad MRK. Free energy calculations shed light on the nuclear pore complex's selective barrier nature. Biophys J. 2021;120(17):3628-40.
33. Lim RY, Aebi U, Fahrenkrog B. Towards reconciling structure and function in the nuclear pore complex. Histochemistry and cell biology. 2008;129(2):105-16.
34. Atkinson CE, Mattheyses AL, Kampmann M, Simon SM. Conserved spatial organization of FG domains in the nuclear pore complex. Biophys J. 2013;104(1):37-50.
35. Sakiyama Y, Mazur A, Kapinos LE, Lim RY. Spatiotemporal dynamics of the nuclear pore complex transport barrier resolved by high-speed atomic force microscopy. Nat Nanotechnol. 2016;11(8):719-23.
36. Peyro M, Soheilypour M, Lee BL, Mofrad MR. Evolutionarily Conserved Sequence Features Regulate the Formation of the FG Network at the Center of the Nuclear Pore Complex. Scientific reports. 2015;5:15795.
37. Huang K, Tagliazucchi M, Park SH, Rabin Y, Szleifer I. Nanocompartmentalization of the Nuclear Pore Lumen. Biophys J. 2020;118(1):219-31.
38. Chowdhury R, Sau A, Musser SM. Super-resolved 3D tracking of cargo transport through nuclear pore complexes. Nat Cell Biol. 2022;24(1):112-22.

39.	Strawn LA, Shen T, Shulga N, Goldfarb DS, Wente SR. Minimal nuclear pore complexes define FG repeat domains essential for transport. Nat Cell Biol. 2004;6(3):197-206.
40.	Bayliss R, Littlewood T, Strawn LA, Wente SR, Stewart M. GLFG and FxFG nucleoporins bind to overlapping sites on importin-beta. J Biol Chem. 2002;277(52):50597-606.
41.	Allen NP, Huang L, Burlingame A, Rexach M. Proteomic analysis of nucleoporin interacting proteins. J Biol Chem. 2001;276(31):29268-74.
42.	Fiserova J, Richards SA, Wente SR, Goldberg MW. Facilitated transport and diffusion take distinct spatial routes through the nuclear pore complex. J Cell Sci. 2010;123(Pt 16):2773-80.
43.	Yang W. Distinct, but not completely separate spatial transport routes in the nuclear pore complex. Nucleus. 2013;4(3):166-75.
44.	Ma J, Goryaynov A, Yang W. Super-resolution 3D tomography of interactions and competition in the nuclear pore complex. Nat Struct Mol Biol. 2016;23(3):239-47.
45.	Davis LK, Ford IJ, Hoogenboom BW. Crowding-induced phase separation of nuclear transport receptors in FG nucleoporin assemblies. Elife. 2022;11.
46.	Gao Y, Skowyra ML, Feng P, Rapoport TA. Protein import into peroxisomes occurs through a nuclear pore-like phase. Science. 2022;378(6625):eadf3971.
47.	Meinema AC, Poolman B, Veenhoff LM. Quantitative analysis of membrane protein transport across the nuclear pore complex. Traffic. 2013;14(5):487-501.
48.	Mudumbi KC, Czapiewski R, Ruba A, Junod SL, Li Y, Luo W, et al. Nucleoplasmic signals promote directed transmembrane protein import simultaneously via multiple channels of nuclear pores. Nat Commun. 2020;11(1):2184.
49.	Kubitscheck U, Siebrasse JP. Kinetics of transport through the nuclear pore complex. Semin Cell Dev Biol. 2017;68:18-26.
50.	Paci G, Zheng T, Caria J, Zilman A, Lemke EA. Molecular determinants of large cargo transport into the nucleus. Elife. 2020;9.
51.	Visa N, Alzhanova-Ericsson AT, Sun X, Kiseleva E, Bjorkroth B, Wurtz T, et al. A pre-mRNA-binding protein accompanies the RNA from the gene through the nuclear pores and into polysomes. Cell. 1996;84(2):253-64.
52.	Zimmerli CE, Allegretti M, Rantos V, Goetz SK, Obarska-Kosinska A, Zagoriy I, et al. Nuclear pores dilate and constrict in cellulo. Science. 2021;374(6573):eabd9776.
53.	Pulupa J, Prior H, Johnson DS, Simon SM. Conformation of the nuclear pore in living cells is modulated by transport state. Elife. 2020;9.
54.	Zheng T, Zilman A. Self-regulation of the nuclear pore complex enables clogging-free crowded transport. 2022.
55.	Jovanovic-Talisman T, Tetenbaum-Novatt J, McKenney AS, Zilman A, Peters R, Rout MP, et al. Artificial nanopores that mimic the transport selectivity of the nuclear pore complex. Nature. 2009;457(7232):1023-7.
56.	Lennon KM, Soheilypour M, Peyro M, Wakefield DL, Choo GE, Mofrad MRK, et al. Characterizing Binding Interactions That Are Essential for Selective Transport through the Nuclear Pore Complex. Int J Mol Sci. 2021;22(19).
57.	Derrer CP, Mancini R, Vallotton P, Huet S, Weis K, Dultz E. The RNA export factor Mex67 functions as a mobile nucleoporin. J Cell Biol. 2019;218(12):3967-76.

Jan 19 '23  16

58.	Ben-Yishay R, Mor A, Shraga A, Ashkenazy-Titelman A, Kinor N, Schwed-Gross A, et al. Imaging within single NPCs reveals NXF1's role in mRNA export on the cytoplasmic side of the pore. J Cell Biol. 2019;218(9):2962-81.
59.	Kalita J, Kapinos LE, Zheng T, Rencurel C, Zilman A, Lim RYH. Karyopherin enrichment and compensation fortifies the nuclear pore complex against nucleocytoplasmic leakage. J Cell Biol. 2022;221(3).
60.	Fragasso A, de Vries HW, Andersson J, van der Sluis EO, van der Giessen E, Onck PR, et al. Transport receptor occupancy in nuclear pore complex mimics. Nano Research. 2022;15(11):9689-703.
61.	Hough LE, Dutta K, Sparks S, Temel DB, Kamal A, Tetenbaum-Novatt J, et al. The molecular mechanism of nuclear transport revealed by atomic-scale measurements. Elife. 2015;4.
62.	Stewart M. Function of the Nuclear Transport Machinery in Maintaining the Distinctive Compositions of the Nucleus and Cytoplasm. Int J Mol Sci. 2022;23(5).
63.	Stewart M. Structural basis for the nuclear protein import cycle. Biochem Soc Trans. 2006;34(Pt 5):701-4.
64.	Raveh B, Karp JM, Sparks S, Dutta K, Rout MP, Sali A, et al. Slide-and-exchange mechanism for rapid and selective transport through the nuclear pore complex. Proc Natl Acad Sci U S A. 2016;113(18):E2489-97.
65.	Milles S, Mercadante D, Aramburu IV, Jensen MR, Banterle N, Koehler C, et al. Plasticity of an ultrafast interaction between nucleoporins and nuclear transport receptors. Cell. 2015;163(3):734-45.
66.	Sakiyama Y, Panatala R, Lim RYH. Structural dynamics of the nuclear pore complex. Semin Cell Dev Biol. 2017;68:27-33.
67.	Hayama R, Sparks S, Hecht LM, Dutta K, Karp JM, Cabana CM, et al. Thermodynamic characterization of the multivalent interactions underlying rapid and selective translocation through the nuclear pore complex. J Biol Chem. 2018;293(12):4555-63.
68.	Najbauer EE, Ng SC, Griesinger C, Gorlich D, Andreas LB. Atomic resolution dynamics of cohesive interactions in phase-separated Nup98 FG domains. Nat Commun. 2022;13(1):1494.
69.	Frey S, Richter RP, Gorlich D. FG-rich repeats of nuclear pore proteins form a three-dimensional meshwork with hydrogel-like properties. Science. 2006;314(5800):815-7.
70.	Ader C, Frey S, Maas W, Schmidt HB, Gorlich D, Baldus M. Amyloid-like interactions within nucleoporin FG hydrogels. Proc Natl Acad Sci U S A. 2010;107(14):6281-5.
71.	Ibanez de Opakua A, Geraets JA, Frieg B, Dienemann C, Savastano A, Rankovic M, et al. Molecular interactions of FG nucleoporin repeats at high resolution. Nat Chem. 2022;14(11):1278-85.
72.	Fragasso A, de Vries HW, Andersson J, van der Sluis EO, van der Giessen E, Dahlin A, et al. A designer FG-Nup that reconstitutes the selective transport barrier of the nuclear pore complex. Nat Commun. 2021;12(1):2010.
73.	Ketterer P, Ananth AN, Laman Trip DS, Mishra A, Bertosin E, Ganji M, et al. DNA origami scaffold for studying intrinsically disordered proteins of the nuclear pore complex. Nat Commun. 2018;9(1):902.
74.	Fisher PDE, Shen Q, Akpinar B, Davis LK, Chung KKH, Baddeley D, et al. A Programmable DNA Origami Platform for Organizing Intrinsically Disordered Nucleoporins within Nanopore Confinement. ACS nano. 2018;12(2):1508-18.

75. Andersson J, Svirelis J, Medin J, Jarlebark J, Hailes R, Dahlin A. Pore performance: artificial nanoscale constructs that mimic the biomolecular transport of the nuclear pore complex. Nanoscale Adv. 2022;4(23):4925-37.
76. Frey S, Gorlich D. A saturated FG-repeat hydrogel can reproduce the permeability properties of nuclear pore complexes. Cell. 2007;130(3):512-23.
77. Halfmann R, Wright JR, Alberti S, Lindquist S, Rexach M. Prion formation by a yeast GLFG nucleoporin. Prion. 2012;6(4):391-9.
78. Guo L, Kim HJ, Wang H, Monaghan J, Freyermuth F, Sung JC, et al. Nuclear-Import Receptors Reverse Aberrant Phase Transitions of RNA-Binding Proteins with Prion-like Domains. Cell. 2018;173(3):677-92 e20.
79. Mittag T, Pappu RV. A conceptual framework for understanding phase separation and addressing open questions and challenges. Molecular cell. 2022;82(12):2201-14.
80. Mattheyses AL, Kampmann M, Atkinson CE, Simon SM. Fluorescence anisotropy reveals order and disorder of protein domains in the nuclear pore complex. Biophys J. 2010;99(6):1706-17.
81. Emilsson G, Xiong K, Sakiyama Y, Malekian B, Ahlberg Gagner V, Schoch RL, et al. Polymer brushes in solid-state nanopores form an impenetrable entropic barrier for proteins. Nanoscale. 2018;10(10):4663-9.
82. Yamada J, Phillips JL, Patel S, Goldfien G, Calestagne-Morelli A, Huang H, et al. A bimodal distribution of two distinct categories of intrinsically disordered structures with separate functions in FG nucleoporins. Mol Cell Proteomics. 2010;9(10):2205-24.
83. Ng SC, Gorlich D. A simple thermodynamic description of phase separation of Nup98 FG domains. Nat Commun. 2022;13(1):6172.
84. Labokha AA, Gradmann S, Frey S, Hulsmann BB, Urlaub H, Baldus M, et al. Systematic analysis of barrier-forming FG hydrogels from Xenopus nuclear pore complexes. Embo J. 2013;32(2):204-18.
85. Frey S, Rees R, Schunemann J, Ng SC, Funfgeld K, Huyton T, et al. Surface Properties Determining Passage Rates of Proteins through Nuclear Pores. Cell. 2018;174(1):202-17 e9.
86. Hulsmann BB, Labokha AA, Gorlich D. The permeability of reconstituted nuclear pores provides direct evidence for the selective phase model. Cell. 2012;150(4):738-51.
87. Hough L, Dutta K, Kamal A, Temel D, Sparks S, Tetenbaum-Novatt J, et al. Atomic Scale Dynamic Behavior of the Nuclear Transport Selectivity Barrier. Molecular biology of the cell. 2014;25.
88. Milles S, Lemke EA. Mapping multivalency and differential affinities within large intrinsically disordered protein complexes with segmental motion analysis. Angew Chem Int Ed Engl. 2014;53(28):7364-7.
89. Sparks S, Temel DB, Rout MP, Cowburn D. Deciphering the "Fuzzy" Interaction of FG Nucleoporins and Transport Factors Using Small-Angle Neutron Scattering. Structure. 2018;26(3):477-84 e4.
90. Lee CF, Brangwynne CP, Gharakhani J, Hyman AA, Julicher F. Spatial organization of the cell cytoplasm by position-dependent phase separation. Phys Rev Lett. 2013;111(8):088101.
91. Hyman AA, Weber CA, Julicher F. Liquid-liquid phase separation in biology. Annual review of cell and developmental biology. 2014;30:39-58.

92. Celetti G, Paci G, Caria J, VanDelinder V, Bachand G, Lemke EA. The liquid state of FG-nucleoporins mimics permeability barrier properties of nuclear pore complexes. J Cell Biol. 2020;219(1).
93. Schmidt HB, Gorlich D. Nup98 FG domains from diverse species spontaneously phase-separate into particles with nuclear pore-like permselectivity. Elife. 2015;4.
94. Schmidt HB, Gorlich D. Transport Selectivity of Nuclear Pores, Phase Separation, and Membraneless Organelles. Trends in biochemical sciences. 2016;41(1):46-61.
95. Taylor N, Elbaum-Garfinkle S, Vaidya N, Zhang H, Stone HA, Brangwynne CP. Biophysical characterization of organelle-based RNA/protein liquid phases using microfluidics. Soft Matter. 2016;12(45):9142-50.
96. Feric M, Vaidya N, Harmon TS, Mitrea DM, Zhu L, Richardson TM, et al. Coexisting Liquid Phases Underlie Nucleolar Subcompartments. Cell. 2016;165(7):1686-97.
97. Yamazaki T, Yamamoto T, Hirose T. Micellization: A new principle in the formation of biomolecular condensates. Front Mol Biosci. 2022;9:974772.
98. Milner ST, Witten TA, Cates ME. Theory of the grafted polymer brush. Macromolecules. 2002;21(8):2610-9.
99. Alberti S, Hyman AA. Biomolecular condensates at the nexus of cellular stress, protein aggregation disease and ageing. Nature reviews Molecular cell biology. 2021;22(3):196-213.
100. Weisbrod S. Active chromatin. Nature. 1982;297(5864):289-95.
101. Jones RG. Compendium of polymer terminology and nomenclature : IUPAC recommendations 2008. 2nd ed. ed. Cambridge, UK :: RSC Pub.; 2009.
102. Michieletto D, Marenda M. Rheology and Viscoelasticity of Proteins and Nucleic Acids Condensates. JACS Au. 2022;2(7):1506-21.
103. Lovestam S, Koh FA, van Knippenberg B, Kotecha A, Murzin AG, Goedert M, et al. Assembly of recombinant tau into filaments identical to those of Alzheimer's disease and chronic traumatic encephalopathy. Elife. 2022;11.
104. Kuiper EFE, Gallardo P, Bergsma T, Mari M, Kolbe Musskopf M, Kuipers J, et al. The chaperone DNAJB6 surveils FG-nucleoporins and is required for interphase nuclear pore complex biogenesis. Nat Cell Biol. 2022;24(11):1584-94.
105. Prophet SM, Rampello AJ, Niescier RF, Gentile JE, Mallik S, Koleske AJ, et al. Atypical nuclear envelope condensates linked to neurological disorders reveal nucleoporin-directed chaperone activities. Nat Cell Biol. 2022;24(11):1630-41.
106. Milles S, Huy Bui K, Koehler C, Eltsov M, Beck M, Lemke EA. Facilitated aggregation of FG nucleoporins under molecular crowding conditions. EMBO reports. 2013;14(2):178-83.
107. Thomas L, Ismail BT, Askjaer P, Seydoux G. 2022.
108. Springhower CE, Rosen MK, Chook YM. Karyopherins and condensates. Curr Opin Cell Biol. 2020;64:112-23.
109. Coyne AN, Rothstein JD. Nuclear pore complexes - a doorway to neural injury in neurodegeneration. Nat Rev Neurol. 2022;18(6):348-62.
110. Li N, Lagier-Tourenne C. Nuclear pores: the gate to neurodegeneration. Nat Neurosci. 2018;21(2):156-8.
111. Spannl S, Tereshchenko M, Mastromarco GJ, Ihn SJ, Lee HO. Biomolecular condensates in neurodegeneration and cancer. Traffic. 2019;20(12):890-911.

112.	Denning DP, Patel SS, Uversky V, Fink AL, Rexach M. Disorder in the nuclear pore complex: the FG repeat regions of nucleoporins are natively unfolded. Proc Natl Acad Sci U S A. 2003;100(5):2450-5.
113.	Yu M, Heidari M, Mikhaleva S, Tan PS, Mingu S, Ruan H, et al. Deciphering the conformations and dynamics of FG-nucleoporins in situ. bioArkiv 2022.
114.	Davis LK, Ford IJ, Saric A, Hoogenboom BW. Intrinsically disordered nuclear pore proteins show ideal-polymer morphologies and dynamics. Phys Rev E. 2020;101(2-1):022420.
115.	Davis LK, Saric A, Hoogenboom BW, Zilman A. Physical modeling of multivalent interactions in the nuclear pore complex. Biophys J. 2021;120(9):1565-77.
116.	Barbato S, Kapinos LE, Rencurel C, Lim RYH. Karyopherin enrichment at the nuclear pore complex attenuates Ran permeability. J Cell Sci. 2020;133(3).
117.	McCarthy MR, Lusk CP. One ring doesn't rule them all: Distinct nuclear pore complexes in a single cell. Cell. 2022;185(2):230-1.
118.	Petrovic S, Samanta D, Perriches T, Bley CJ, Thierbach K, Brown B, et al. Architecture of the linker-scaffold in the nuclear pore. Science. 2022;376(6598):eabm9798.
119.	Ghavami A, Van der Giessen E, Onck PR. Sol-gel transition in solutions of FG-Nups of the nuclear pore complex. Extreme Mech Lett. 2018;22:36-41.
120.	Gu C, Vovk A, Zheng T, Coalson RD, Zilman A. The Role of Cohesiveness in the Permeability of the Spatial Assemblies of FG Nucleoporins. Biophys J. 2019;116(7):1204-15.
121.	Pulupa J, Rachh M, Tomasini MD, Mincer JS, Simon SM. A coarse-grained computational model of the nuclear pore complex predicts Phe-Gly nucleoporin dynamics. J Gen Physiol. 2017;149(10):951-66.
122.	Winogradoff D, Chou HY, Maffeo C, Aksimentiev A. Percolation transition prescribes protein size-specific barrier to passive transport through the nuclear pore complex. Nat Commun. 2022;13(1):5138.
123.	Ghavami A, Veenhoff LM, van der Giessen E, Onck PR. Probing the disordered domain of the nuclear pore complex through coarse-grained molecular dynamics simulations. Biophys J. 2014;107(6):1393-402.
124.	Soheilypour M, Mofrad MRK. Agent-Based Modeling in Molecular Systems Biology. BioEssays : news and reviews in molecular, cellular and developmental biology. 2018;40(7):e1800020.
125.	Yang W, Musser SM. Nuclear import time and transport efficiency depend on importin beta concentration. J Cell Biol. 2006;174(7):951-61.
126.	Kuhn TS. The structure of scientific revolutions: Chicago University of Chicago Press; 1970.
127.	Beyer DK, Forero A. Mechanisms of Antiviral Immune Evasion of SARS-CoV-2. J Mol Biol. 2022;434(6):167265.
128.	Zhang K, Miorin L, Makio T, Dehghan I, Gao S, Xie Y, et al. Nsp1 protein of SARS-CoV-2 disrupts the mRNA export machinery to inhibit host gene expression. Sci Adv. 2021;7(6):eabe7386.